\documentstyle[12pt]{article}

\def\beq{\begin{equation}}
\def\eeq{\end{equation}}
\def\bea{\begin{eqnarray}}
\def\eea{\end{eqnarray}}

\begin{document}
\pagestyle{empty}
\begin{flushright}
{CERN-TH/97-205}\\
{OUTP-98-21P}\\ 
\end{flushright}
\vspace*{5mm}

\begin{center}

{\large {\bf Hierarchies of $R$-violating Interactions \\from Family
Symmetries}}


\vspace*{1cm} {\bf John Ellis} and {\bf Smaragda Lola}\\[0pt]
\vspace{0.3cm} Theoretical Physics Division, CERN \\[0pt]
1211 Geneva 23, Switzerland\\[0pt]
and \\[0.4pt]
{\bf Graham G. Ross} \\[0pt]
\vspace{0.3cm} Department of Theoretical Physics,\\[0pt]
University of Oxford,\\[0pt]
Oxford, United Kingdom\\[0pt]

\vspace*{1.7cm} {\bf ABSTRACT} \\[0pt]

\end{center}


\noindent We investigate the possibility of constructing models of $R$%
-violating $LQ\bar{D}$ Yukawa couplings using a single $U(1)$
flavour-symmetry group and supermultiplet charge assignments that are
compatible with the known hierarchies of quark and lepton masses. 
The mismatch of mass and current eigenstates inferred from the
known charged-current mixing induces the propagation of strong
phenomenological constraints on some $R$-violating couplings to many
others. Applying these constraints,
we look for flavour-symmetry models that are consistent with
different
squark-production hypotheses devised to explain the possible HERA large-$
Q^{2}$ anomaly. The 
$e^{+}d \rightarrow {\tilde{t}}$ 
interpretation of the HERA data is accommodated relatively
easily, at the price of postulating an extra
parity. 
The $e^{+}s\rightarrow {\tilde{t}}$ interpretation of
the events requires models to have only
small (2,3) mixing in the down quark sector.
The $ e^{+} d \rightarrow {\tilde{c}}$ mechanism cannot
be accommodated without large violations of 
squark-mass universality,
due to the very strong experimental constraints
on $R$-violating operators.
We display a model in which
baryon decay due to dangerous dimension-five
operators is automatically suppressed.


\begin{flushleft}
CERN-TH/97-205\\
{OUTP-98-21P}\\ 
March  1998
\end{flushleft}
\vfill\eject


\setcounter{page}{1} \pagestyle{plain}

\section{Introduction}

Although the minimal supersymmetric Standard Model (MSSM)
has dominated the
phenomenological studies of supersymmetric signals
\cite{HabKan}, it has long been known
that the symmetries of the Standard Model allow additional dimension-four
couplings which may lead to interesting baryon- and lepton-
number-violating processes~\cite{Rpar}.
These couplings are expected to
be present in the low energy Lagrangian, unless
forbidden by a symmetry such as $R$ parity~\cite{fayet}.
The complete set of such terms in the
superpotential is:

\[
\lambda L_{i}L_{j}{\bar{E}}_{k}+\lambda ^{\prime }L_{i}Q_{j}{\bar{D}_{k}}%
+\lambda ^{\prime \prime }{\bar{U}_{i}}{\bar{D}_{j}}{\bar{D}_{k}} 
\]

where the $L(Q)$ are the left-handed lepton (quark) superfields, and the
${\bar{E}}$,(${\bar{D}},{\bar{U}}$) are the corresponding right-handed fields. 
The
symmetries of the model imply that there are 45 operators in total.
However,
there are many experimental
constraints on these operators and their combinations, of which the most
stringent
comes from proton stability and excludes the simultaneous presence of
certain products of $LQ\bar{D}$ and $\bar{U}\bar{D}\bar{D}$ 
couplings~\cite{SMVIS}. In addition, experimental constraints from the
non-observation of modifications to Standard Model processes, or of possible
exotic processes, gives bounds for most of the
operators~\cite{constraints} and some combinations involving pairs of
fermion generations.
On the other hand, possible
strong limits on $R$-parity violating interactions from
cosmological arguments~\cite{CDEO} can be avoided
in various schemes \cite{DRR},
including the case of electroweak baryogenesis 
~\cite{Barel}.

The large number of $R$-violating couplings complicates the 
systematic discussion of
the phenomenological implications of these constraints. To date, most
phenomenological
analyses have assumed the dominance of a single operator, arguing that the
Yukawa couplings of the Standard model display just such a property. In
flavour-symmetry models, the
dominant operator is naturally specified in the
quark and lepton {\it current} basis. It is
plausible to assume that mass mixing will induce non-zero
coefficients for operators related to the dominant one.
In Section 2 of this paper, we
pursue this argument and compile the corresponding implications of
some severe upper limits on particular $R$-violating interactions.

There have been many attempts to understand quark and lepton
masses and the mixing angles between mass and current eigenstates
using models for family symmetries~\cite{famsym}.
Some of these reproduce
successfully the qualitative features of fermion masses and mixings,
and so provide plausible frameworks for analyzing the 
possible hierarchy of $R$-violating interactions
\cite{MODELS}. In
this paper we consider models based on a single $U(1)$ family
symmetry, with fermion charges constrained by the observed hierarchy of
fermion masses and mixing angles \cite{IR}.
Such models are discussed in
Section 3, where problems arising from symmetric mass matrices
and from constraints on products of operators are emphasized.

As a specific application of this
analysis, we look in Section 4 for models that might accommodate
the proposed $R$-violating
interpretations \cite{Rviol}
of the possible HERA large-Q$^{2}$ anomaly \cite{H1,ZEUS}.
Of the $45$ operators mentioned earlier,
$9$ could in principle lead to resonant squark production at HERA. 
Of these, only the $\lambda'_{121}, \lambda'_{131}$ and $\lambda'_{132}$
cases survived an initial
confrontation with other experimental constraints~\cite{Rviol}
The suggestion that the apparent HERA excess
may be due to single sparticle production via
some $R$-violating couplings 
may not be gaining support~\cite{Jer}.
Nevertheless, our analysis gives an indication which of the proposed
mechanisms may be compatible with $U(1)$ family-symmetry models.
Within the framework of the most symmetric schemes describing fermion masses,
we find the bounds on R-violating couplings are so strong
that such schemes do not lead to a significant excess of 
HERA events over the Standard Model prediction. However, in more 
general schemes we find that the $\lambda'_{131}$ interpretation is easy to
accommodate, whereas
the $\lambda'_{121}$ interpretation has 
difficulties with squark mass universality.
We indicate how to construct a model consistent with
the $\lambda'_{132}$ interpretation,
though we do not present a specific example. 
We also
show how the structure of the quark
and lepton mass matrices would be strongly constrained by the
confirmation of such an $R$-violating signal.

\section{$R$ Violation and Family Symmetries}

In order to obtain a realistic form for the quark and lepton masses and
mixing angles, it is necessary to have non-diagonal forms for the mass
matrices in the current basis. Diagonalising the mass matrix then implies
that
the mass eigenstates are mixtures of the current eigenstates. Attempts to
make sense of the pattern of fermion 
masses and mixing angles often start with a
family symmetry in the current basis which, when exact, allows only the
third generation of quarks and leptons to acquire mass. Spontaneous breaking
of this symmetry then allows other entries of the mass matrix to be
non-zero. If the
breaking is weak, these entries will be small, offering an explanation for
the observed hierarchy of fermion masses and mixing angles. 

If $R$ parity is violated, such a
symmetry would have
important implications for $R$-violating operators, since couplings
with
different family structures would also appear with different powers of the
family symmetry-breaking parameter. This is consistent with the common
assumption that a single $R$-violating operator dominates. However,
this assumption would apply in the current quark and lepton basis, and in
the mass-eigenstate
basis there would be several operators corresponding to the original
dominant one in the current basis. Any given family-symmetry model
would make characteristic predictions for the pattern of these related
operators. Since there are stringent bounds on some of $R$-violating
operators, particularly on those involving the first family and on
some combinations that mix families, an analysis
of such sub-leading operators in the mass-eigenstate basis
may provide the most stringent bounds on the operators
related by mass mixing. In addition, there could
also be further contributions due to operators that are sub-dominant
in the current basis, with strengths given by powers of the
family symmetry-breaking parameter that are calculable in any
given model. 

The relation between the forms of the mass matrix in the current and the
mass eigenstate basis is given by 

\begin{eqnarray}
M_{u}^{\prime } &=&V_{u}^{L}.M_{u}^{Diag}.(V_{u}^{R})^{\dagger }  \nonumber
\\
M_{d}^{\prime } &=&V_{d}^{L}.M_{d}^{Diag}.(V_{d}^{R})^{\dagger }  \nonumber
\\
M_{\ell }^{\prime } &=&V_{\ell }^{L}.M_{\ell }^{Diag}.(V_{\ell
}^{R})^{\dagger }
\end{eqnarray}

where $V_{u,d,\ell}^{L,R}$ are the unitary matrices relating the 
left- and right-handed $u,$ $d$ and $ \ell $ 
current eigenstates to their mass
eigenstates. We use the notation
${\cal L}_{mass} = \bar{\Psi}'_L M' \Psi'_R$ for all mass terms, so
that the $V_L$ are given by diagonalising 
$M' M^{'\dagger}$, whilst the 
$V_R$ are obtained by diagonalising 
$M^{'\dagger} M'$. Only information on the
entries of the Cabibbo-Kobayashi-Maskawa (CKM) product matrix 

\begin{equation}
V^{CKM}=V_{u}^{L\dagger }V_{d}^{L}
\end{equation}

is provided by experiments to date.

In general, one can construct models where the quark mixing is either in the
up sector, or in the down sector, or both. In the class of models
studied in this paper,
in which the mass matrices have small off-diagonal
entries generated by spontaneous breaking of a family symmetry,
one may obtain useful connections between the mixing matrices and the
elements of the mass matrices in the current basis by perturbative
expressions
for the off-diagonal elements, which are given in the Appendix. From this
general analysis, it may be seen that in the specific case
of the CKM mixing matrix the leading-order contribution
comes from the $d$-quark mass-matrix elements 
that lie above the diagonal in our representation. As a
result, we have little experimental input
to guide us in constructing models for the elements below the
diagonal. However, it has has been noted for some time that a
phenomenologically
successful relationship results if one assumes a ``texture zero'' in the
(1,1) position {\it and} symmetry between the (1,2) and (2,1) matrix
elements~\cite{RRR}.
In this case one finds the relation 

\[
\mid V_{ud}\mid =(\frac{m_{d}}{m_{s}}+\frac{m_{u}}{m_{c}}+2\sqrt{\frac{%
m_{d}m_{u}}{m_{s}m_{c}}}\cos {\phi })^{\frac{1}{2}} 
\]

where $\phi $ is the usual CP-violating phase in the CKM matrix. The
fact
that this relation works well is the only phenomenological indication we
have for a symmetric structure of the mass matrices, and it may
just be accidental. Nevertheless, we think it a useful starting
point for our analysis, so we consider first a simple model capable of
yielding
this form and accommodating the remaining fermion masses and mixing
angles \cite{IR}.

\begin{table}[h]
\centering
\begin{tabular}{|c|ccccccc|}
\hline
& $Q_i$ & $\bar{U}_i$ & $\bar{D}_i$ & $L_i$ & $\bar{E}_i$ & $H_2$ & $H_1$ \\ 
\hline
$U(1)$ & $a _i$ & $a _i$ & $a _i$ & $b_i$ & $b_i$ & $-2a _3$ & $w a _3$
\\ \hline
\end{tabular}
\caption{{\it Assignments of $U(1)$ charges.}}
\end{table}

The model consists of a single $U(1)$ family symmetry with the same charges
for the left- and right-handed states, as shown in Table 1,
where, e.g.,
the choice $a_{i}=(-4,1,0)$ gives an acceptable pattern for the mass
matrices.
Suppressing unknown numerical factors and phases, which are all expected to
be of order unity, with these charge assignments the up-quark mass matrix
takes the form 

\[
M^{up}=\left( 
\begin{array}{ccc}
\epsilon ^{8} & \epsilon ^{3} & \epsilon ^{4} \\ 
\epsilon ^{3} & \epsilon ^{2} & \epsilon \\ 
\epsilon ^{4} & \epsilon & 1
\end{array}
\right) 
\label{mm}
\]

The down-quark mass matrix has a similar form, but with a different
expansion parameter $\bar{\epsilon}\approx \sqrt{\epsilon }$. Since the up
and down sectors have similar structures, mixing is present in both sectors,
though it may be larger in the down sector, simply because $\bar{\epsilon}
>\epsilon $. 
For the mass matrices of~\cite{IR} that 
we consider initially,
one finds the following
expressions for the quark mixing matrices~\footnote{
Lepton mixing is discussed in the next section.}: 

\[
V_{u}^{L,R}\approx \left( 
\begin{array}{ccc}
1 & \epsilon & 2\epsilon ^{4} \\ 
-\epsilon & 1 & \epsilon \\ 
\epsilon ^{2} & -\epsilon & 1
\end{array}
\right) ,\;\;V_{d}^{L,R}\approx \left( 
\begin{array}{ccc}
1 & \bar{\epsilon} & 2\bar{\epsilon}^{4} \\ 
-\bar{\epsilon} & 1 & \bar{\epsilon} \\ 
\bar{\epsilon}^{2} & -\bar{\epsilon} & 1
\end{array}
\right) 
\]

using the second-order
perturbation-theory formulae given in the Appendix.

We now discuss the importance of this mixing for
$R$-parity violation. The most relevant experimental constraints are
those on the
operators $L_{1}Q_{1}\bar{D}_{1}$ and $L_{1}Q_{3}\bar{D}_{3}$, for which
$\lambda _{111}^{\prime }\leq 0.002$ from nuclear $\beta \beta $ decay 
\cite{HirVer} and for squark and gluino masses of 200 GeV,
while 
$\lambda _{133}^{\prime }\leq 0.001$ from bounds~\cite{numass}
on Majorana neutrino masses, again assuming masses of
200~GeV for the sparticles.~\footnote{The quoted bound is
clearly
only approximate, as the exact value depends on soft
parameters~\cite{JOSI}.} 
However, operators related to these operators by mass mixing
are also strongly constrained by these bounds. Consider first the
relations 
\begin{equation}
\begin{array}{l}
(L_{1}Q_{1}\bar{D}_{1})^{\prime }=L_{1}Q_{1}\bar{D}_{1}+\epsilon L_{1}Q_{2}%
\bar{D}_{1}+2\epsilon ^{4}L_{1}Q_{3}\bar{D}_{1}+... \\ 
(L_{1}Q_{3}\bar{D}_{3})^{\prime }=L_{1}Q_{3}\bar{D}_{3}-\bar{\epsilon}%
L_{1}Q_{3}\bar{D}_{2}+\bar{\epsilon}^{2}L_{1}Q_{3}\bar{D}_{1}+...
\end{array}
\label{opmix}
\end{equation}

where the notation $()^{\prime }$ denotes effective operators in a
current-eigenstate basis. We see that the operators $L_{1}Q_{2}
\bar{D}_{1}$ and $L_{1}Q_{3}\bar{D}_{1}$ mix with $L_{1}Q_{1}\bar{D}_{1}$,
so their coefficients are constrained to be the appropriate mixing
coefficient ($\epsilon ^{-1}$ and $(1/ (2 \epsilon^4)$, respectively)
times
the bound on $\lambda _{111}^{\prime }$. Similarly, the coefficients of
the operators $L_{1}Q_{3}\bar{D}_{2}$ and $L_{1}Q_{3}\bar{D}_{1}$ are
constrained to be less than $\bar{\epsilon}^{-1}$ and $\bar{\epsilon}^{-2}$
times the bound on $\lambda _{133}^{\prime }$, respectively.

We display below, as an example,
matrices of upper limits 
on $L_1 Q_j \bar{D}_k$ operators.
These limits follow from the
mixing in this particular model, combined with
the experimental upper bounds for sfermion
masses of 200 GeV~\cite{constraints,HirVer,numass,noi,Agashe,atpar}.
We first look at the bounds that arise from mixing with the
$\lambda'_{111}$ operator, tabulating the direct experimental bounds
in cases where they are stronger than those originating from the
$\lambda'_{111}$ mixing:
\[
L_{1jk}^{({\rm from} \;
111)} < \left( 
\begin{array}{ccc}
0.002 & 0.009 & 0.04 \\
0.038 (0.009) & 0.03 & 0.3 \\
0.07 & 0.56 & 0.001
\end{array}
\right)
\]

In certain entries we have two values,
because  bounds on
$e u \bar{d}$  terms
involve mixing in the up-quark sector,
whereas  bounds on
$\nu_e d \bar{d}$ terms
involve mixing in the down-quark sector.
Then we repeat the analysis for mixing with the
$\lambda'_{133}$ coupling:

\[
L_{1jk}^{(
{\rm from} \;
133)} < \left( 
\begin{array}{ccc}
0.002 & 0.04 & 0.04 (0.02) \\
0.07 & 0.03 (0.02) & 0.02 (0.004) \\
0.02 & 0.004 & 0.001
\end{array}
\right)
\]

and finally we gather all the best limits for the matrix elements in
this particular model:

\[
L_{1jk}^{best} < \left( 
\begin{array}{ccc}
0.002 & 0.009 & 0.04 (0.02) \\
0.02 (0.009) & 0.03 (0.02) & 0.02 (0.004) \\
0.02 & 0.004 & 0.001
\end{array}
\right)
\]

Here, the bound  on the (2,1) entry arises 
from constraints on $\lambda'_{121}$ from
$K \rightarrow \pi \nu \bar{\nu}$
\cite{Agashe}, in the case that
$V^{CKM}_{12,21}$ arises predominantly from
the down-quark sector.
At this stage we have not yet taken into account other bounds,
especially bounds on products of $R$-violating couplings that
pose even stricter constraints
\cite{DEB,othpro}. As an example
for this Ansatz, the couplings
$L_{1}Q_2\bar{D}_1$  and 
$L_{1}Q_1\bar{D}_2$  
appear at such an order in the family-symmetry breaking
that the strong bound on the product of these couplings
from contributions to
$\Delta m_K$~\cite{DEB} is not satisfied. We shall return to
this and related issues at a later stage.

It should be noted that the model of~\cite{IR} has
mixing in both
the up and down sectors. Indeed, the (2,3) entry of the down mass matrix is
$\bar{\epsilon}=0.23$, which is much larger than $V_{23}^{CKM}$. Thus, 
to obtain
viable mass matrices in this example, one needs a suppression of the mixing
in $\mid V_{cb}\mid =(a^{\prime }\frac{m_{s}}{m_{b}}+a\frac{m_{c}}{m_{t}}+2%
\sqrt{aa^{\prime }\frac{m_{s}m_{c}}
{m_{b}m_{t}}}\cos {\phi })^{\frac{1}{2}}$~\cite{IR}. This case, 
in which the mixing between states is much larger than
would
have been estimated just using the appropriate CKM mixing matrix element,
serves as a healthy reminder of the potential importance of the
details of the underlying model for fermion
masses when drawing implications for $R$-violating phenomena.

\section{Exploring Hierarchies of $R$-Violating Interactions}

We now consider the effect of the $U(1)$ symmetry
on the pattern of allowed $R$-violating interactions~\cite{MODELS}.
We first recall that possible sets
of quark and lepton charges leading to correct
mass hierarchies are given~\cite{IR} by: \\
{\bf Case 1}: $a_i = b_i = (-4,1,0)$,
where $a_i$ and $b_i$ are the quark and 
lepton charges respectively, and \\ {\bf Case 2}:
$a_i =(-4,1,0)$,
$b_i = (-\frac{7}{2},\frac{1}{2},0)$.

In {\bf Case 1}, where leptons and quarks have the
same charges,
one needs an additional symmetry in order 
to eliminate dimension-four nucleon-decay operators.
This may be done simply by imposing an anomaly-free
flavour-independent
baryon parity \cite{GrL}, under which the fields transform as
\beq
Z_3: (Q,\bar{U},\bar{D},L,\bar{E},H_1,H_2) 
\rightarrow (1,a^2,a,a^2,a^2,a^2,a)
\eeq
This allows only the lepton-number-violating
operators, while forbidding baryon-number-violating ones 
~\footnote{
A flavour-dependent generalisation of this
symmetry has been discussed in~\cite{LR}. In this
case, consistent solutions were found containing
only a subclass of operators violating lepton number ($LL{\bar E}$) and
baryon-number (${\bar U}{\bar D} {\bar D}$).
In this way, it was possible to have 
both lepton and baryon number violation without
disturbing proton stability. However, we do not pursue such models here.}.

In {\bf Case 2}, which is motivated by constaints
on HERA-friendly models,
the lepton charges of the first two
generations are half-integers. On might at first think that
the residual $\tilde{Z}_2$ 
symmetry of the $U(1)$ forbids the $L_{1,2}Q\bar{D}$ operators.
However, it is straightforward to
combine this $\tilde{Z}_2$
with a normal $Z_2^M$ matter parity, so as to allow
these terms while also forbidding the
$\bar{U}\bar{D}\bar{D}$ terms.
This is possible if 
$\tilde{Z}_2 \times Z_2^M$ 
is broken to a residual  $Z_2$
by a field $\Phi$ that is odd under both symmetries.
In this case, $\bar{U}\bar{D}\bar{D}$ is forbidden,
because  it transforms as
$(+,-)$ under
$\tilde{Z}_2 \times Z_2^M$.
Similarly, $L_{1,2}Q\bar{D}$ transforms
as $(-,-)$ and is also forbidden, but it occurs at
${\cal O}\left (\frac{\Phi}{M} \right )$
through the term $(\Phi L Q \bar{D})/M$.

Let us now pass to the charges of the $R$-violating operators.
The first thing to notice is that the
form of the mass matrices only determines the
relative charges of the operators, not their
absolute charges. To see this, note that
that the symmetric structure of $M^{up}$
is unchanged if we add a family-independent 
constant to the charges of the $\bar{U}$ fields.
This shows that, as we have already mentioned,
the charge normalization of our operators
is undetermined by the mass structure.

However, anomaly cancellation must be imposed. With
the general charge assignment given in the first
line of Table~\ref{table:2}, the coefficients of the
$SU(3)^2 \times U(1)$, $SU(2)^2 \times U(1)$ and $U_Y(1) \times U(1)$
anomalies are proportional to $A_{3,2,1}$, where
\bea
A_3 & = & 2\sum a_i + \frac{3}{2} w_1 + \frac{3}{2} w_2 \nonumber \\
A_2 & = & \frac{3}{2} \sum a_i + \frac{1}{2} \sum b_i +
\frac{1}{2} a_3 (w-2) \nonumber \\
A_1 & = & \frac{11}{6} \sum a_i + \frac{3}{2}
\sum b_i + \frac{1}{2} a_3 (w-2) +
4 w_1 + w_2 + 3w_3
\eea
We demand that these should vanish up to a
Green-Schwarz term~\cite{GS}, i.e.,
$A_3:A_2:A_1 = 1:1:5/3$. The effect of
this is shown in Table \ref{table:2}: in the first
row we have a generic charge assignment where the
flavour-independent pieces $w_i$ are to be chosen
such that the anomaly cancellation conditions are
satisfied.
Imposing these conditions and reabsorbing
$a_3$ in the definitions of the charges,
one obtains  the charges shown in the second row
of Table~\ref{table:2}, where
$a'_i \equiv a_i - a_3$, and $b'_i \equiv b_i -b_3$ are of the same
form as discussed above, i.e., $a'_i = (-4,1,0)$, $b'_i = (-4,1,0)$
in Case 1 and $b'_i = (-\frac{7}{2},\frac{1}{2},0)$ in Case 2.

\begin{table}[h]
\centering
\begin{tabular}{|c |ccccccc|}\hline
   &$ Q_i$ & $\bar{U}_i$ &$ \bar{D}_i$ &$ L_i$ & $\bar{E}_i$ &
$H_2$ &
$ H_1$   \\
\hline
  $U(1)$ & $a _i$ & $a _i + w_1$ & $a _i + w_2$  
& $b_i$ & $b_i+w_3 $ & $-2a _3$ &  $ w a _3$ \\
\hline
$U(1)$ & $a'_i$ & $a'_i + w_1$ & $a'_i - w_1$  
& $b'_i$ & $b'_i-w_1 $ & $-w_1$ &  $ w_1$ \\
\hline
\end{tabular}
\caption{{\it Assignments of flavour symmetry charges,
before and after imposing anomaly cancellation.}}
\label{table:2}
\end{table}

We now discuss the possible hierarchies of $R$-violating operators in
the two cases. \\
{\bf Case 1}: In this case, the charges of the operators $O_{ijk} \equiv
L_iL_j\bar{E}_k$ and 
$L_iQ_j\bar{D}_k$ are the same, and depend 
only on the values of $i,j,k$, and not on their order, as
given in Table \ref{table:3}.
We note that the constraints on the
operators $L_1Q_1\bar{D}_1$ from nuclear $\beta\beta$ decay
and on $L_1L_3\bar{E}_3$ 
from bounds on Majorana neutrino masses constrain
the choice of the charge $w_1$. Since the
exact constraint depends on the magnitude of the
expansion parameter for the $R$-violating
couplings,
we need to consider what the constraints are
on this expansion parameter.

\begin{table}[h]

\centering
\begin{tabular}{|c |ccccc|}\hline
ijk   & { 111} & { 121} &{ 122} 
& { 222} & { 131} \\ \hline 
$U(1)$ & $-12-w_1$ &  $-7-w_1$ &  $-2-w_1$ &  $3-w_1$ & 
$-8-w_1$ \\ \hline \hline
ijk & { 133} & { 333} & 
{ 223} & { 233} & { 123} \\
\hline
$U(1)$ &  $-4-w_1$ & $-w_1$ & $2-w_1$ & 
$1-w_1$ & $-3-w_1$ \\
\hline
\end{tabular}
\caption{{\it Operator charges in Case 1.}}
\label{table:3}
\end{table}

In the
case of the mass matrices, it was suggested in~\cite{IR} that the mixing
between Higgs fields carrying different $U(1)$ quantum numbers was
responsible for filling in the remaining elements of the mass matrix. In
this case the expansion parameters $\epsilon$ and $\bar{\epsilon}$ are as
given in~\cite{IR}, with $M_2$, $M_1$ being the mass scales of the
heavy Higgs fields $H_2$, $H_1$ that mix with the light Higgses responsible
for electroweak breaking. The scales of the vacuum expectation values
$<\theta>$,$<\bar{\theta}>$
are bounded from below by $(1 / \sqrt{192}\pi) M_{string}$, the scale
of the $U(1)$ symmetry breaking. Hence
$M_2$ and $M_1$ are bounded from below by 
$\epsilon^{-1}\theta$ and $\bar{\epsilon}^{-1}\theta$, respectively.
In the case of $R$ violation,
mixing
between the operators $LL\bar{E}$ or $LQ\bar{D}$ 
proceeds through heavy lepton or heavy quark mixing
rather than through heavy Higgs exchange.
If the former are
much heavier than the Higgs states, the corresponding expansion parameter $%
\epsilon^{\prime}$ will be much smaller. The limiting case occurs when they
have string-scale masses, corresponding to 

\begin{equation}
\epsilon^{\prime}= \epsilon \frac{M_2}{M_{string}} \geq \frac{<\theta>}{%
M_{string}} = \frac{1}{\sqrt{192} \pi} \approx 0.02 \label{biggest}
\end{equation}

Taking this lower limit for the expansion parameter and using
the constraint
$\lambda _{111}^{\prime }\leq 0.002$ from nuclear $\beta \beta $ decay,
we find that
$|-12-w_1| \geq 2$, whilst the constraint 
$\lambda _{133}^{\prime }\leq 0.001$ from bounds~\cite{numass}
on Majorana neutrino masses
indicates that
$|-4-w_1| \geq 2$.

Next, we note that the magnitudes of the couplings in Table~\ref{table:3}
are symmetric in the three indices $ijk$. This implies, for
example, that at this level the $\lambda'_{121}$ and $\lambda'_{112}$
couplings
should have similar magnitudes.
This must be made consistent with
with the constraint
$(L_1 Q_{2} \bar{D}_1).(L_1 Q_{1} \bar{D}_2)
\leq 4 \cdot 10^{-9}$, which arises from
bounds on $\Delta m_K$~\cite{DEB}.
In the present context, this constraint
indicates that
the relevant charge  $|-7-w_1|$ has to be large, and we
reach our first HERA-unfriendly conclusion:
in this case the $e^+d \rightarrow {\tilde c}$
interpretation of the HERA data would become untenable.

Thirdly, a related problem is that some couplings to muons would
have comparable magnitudes to those listed in Table~\ref{table:3}.
For example, the magnitude of the $\lambda'_{211}$ coupling would be
comparable to that of the $\lambda'_{121}$ coupling.
However, certain products of couplings
involving electrons and muons have to be extremely 
suppressed~\cite{constraints}.
For 200 GeV sfermions,
\bea
\lambda_{231} \lambda_{131} & \leq & 2.8 \cdot  10^{-6}  \nonumber \\
\lambda'_{1k1} \lambda'_{2k2} & \leq & 3.2 \cdot  10^{-6} \nonumber \\
\lambda'_{1k1} \lambda'_{2k1} & \leq & 2 \cdot 10^{-7} \nonumber \\
\lambda'_{11j} \lambda'_{21j} & \leq & 2 \cdot  10^{-7} 
\label{muonelectron}
\eea

Using the form of the mixing matrices for Case 1:
\[
V_{\ell }^{L,R}\approx \left( 
\begin{array}{ccc}
1 & \bar{\epsilon}/3 & 2\bar{\epsilon}^{4} \\ 
-\bar{\epsilon}/3 & 1 & \bar{\epsilon} \\ 
\bar{\epsilon}^{2} & -\bar{\epsilon} & 1
\end{array}
\right) 
\]
we shall see later that
these bounds are so severe as to rule out any possible HERA-friendly model
of this simple type.
Note that, in order to obtain correct 
lepton masses within this 
Ansatz, a factor of $\sim 3$ is needed in the (22) 
element of the mass matrix, and 
this factor also
enters in the mixings.

Other strong constraints on products of couplings are the following:
\bea
\lambda_{1j1} \lambda_{1j2} & \leq & 2.8 \cdot 10^{-6}  \nonumber \\
\lambda'_{i13} \lambda'_{i31} & \leq & 3.2 \cdot  10^{-7} \nonumber \\
\lambda'_{i12} \lambda'_{i21} & \leq & 4 \cdot  10^{-9} ,
\label{others}
\eea
which are particularly stringent in the model
under consideration.
Using these bounds, one finds
\bea
\lambda'_{i13} \leq 6 \cdot 10^{-4} \nonumber \\
\lambda'_{i12} \leq 6 \cdot 10^{-5} ,
\label{indi}
\eea
together with corresponding bounds for
permutations of the indices.

If the reported apparent excess of HERA events at large $Q^2$
were due to production of a single squark by an $R$-violating coupling,
one would need
\bea
\lambda'_{121,131} \approx & 0.04/\sqrt{\cal B} \nonumber \\
\lambda'_{132} \approx  & 0.3/\sqrt{\cal B} \nonumber 
\eea
where ${\cal B}$ is the branching ratio of the decay
$\tilde{q} \rightarrow e^+ q$. 
We see immediately from (\ref{indi})
that the first two possibilities
cannot be realised in this model.
Moreover, we see from (\ref{opmix}) that
$\lambda'_{132} \approx \lambda'_{133}/\bar{\epsilon}
\leq 0.004$, and infer the third possibility cannot 
be realised either.

What happens to the remaining couplings? From the mixing discussed above, 
we have: \\
(i) $\lambda'_{111} = \lambda'_{112}/\bar{\epsilon} \leq
3 \cdot 10^{-4}$, which is a stronger bound than the 
one from neutrinoless $\beta \beta$ decay,~\footnote{We
use the down-quark mixing parameter, which is larger, since the
operator we compare with differs only in the index of $\bar{D}$.} \\
(ii) $\lambda'_{222} = \lambda'_{221}/\bar{\epsilon} \leq
3 \cdot 10^{-4}$, \\
(iii) $\lambda'_{223} = \lambda'_{213}/\bar{\epsilon} 
      = \lambda'_{312}/\bar{\epsilon} \leq
3 \cdot 10^{-4}$ 
or 
$\lambda'_{223} = \lambda'_{213}/ \epsilon
      = \lambda'_{312}/\epsilon \leq 0.0013$. \\
The constraints here are very strict because the
experimental bound on $\lambda'_{312}$ is more severe than
that on $\lambda'_{213}$: since we require these two 
terms to have the same charge, we must
take the stricter limit. We also have \\
(iv) $\lambda'_{233} =  \lambda'_{223}/\bar{\epsilon} \leq 0.0013
(0.006)$, for expansion parameters $\bar{\epsilon}$ and
$\epsilon$ respectively, \\
(v) $\lambda'_{333} =  \lambda'_{233}/\bar{\epsilon}
\leq 0.006 (0.025)$.

In each of these cases, $R$ violation
may be manifest in hadron-hadron colliders.
For sfermion masses of 100 GeV and
$\lambda \geq 10^{-6}$, the lightest supersymmetric
particle is expected
to decay inside the accelerator.
The above constraints allow couplings that are significantly larger
than this lower bound.
Through  a suitable choice 
of $w_1$, the couplings that are
more severely constrained
can be made small, while some others 
can be of importance for collider physics,
though none can be very large in this type of model
with symmetric mass matrices. Hence,
single-squark production via an $R$-violating coupling
is suppressed, and the best signal would be
squark-pair
production followed by $R$-violating decay.

{\bf Case 2}: \\
In this case, the charges of the operators
depend on the flavour-symmetry charge
of the singlet field $\Phi$  that we have introduced.
This does not affect the relative magnitudes
of the $R$-violating couplings, since $\Phi$
appears in all terms. However, 
this charge and the vacuum expectation value 
of $\Phi$ do provide a possible source
of suppression for the $R$-violating couplings.
We take as an indicative value
$a_{\Phi} = 1/2$: the corresponding subclasses of
$LL\bar{E}\Phi$ and
$LQ\bar{D}\Phi$ operators with integer flavour
charge appear in Tables~\ref{table:4} and 
\ref{table:5}. 

\begin{table}[h]
\centering
\begin{tabular}{|c |cccc|}\hline
ijk$(LL\bar{E})$   &$ 121 $ & $122$ &$ 133$ &$ 233$ 
\\ \hline 
$U(1)$ & $-6-w_1$ &  $-2-w_1$ &  $-3-w_1$ &  $1-w_1$ \\
\hline
\end{tabular}
\caption{{\it Integer $LL\bar{E}$ charges for Case 2.}}
\label{table:4}

\vspace*{0.8 cm}

\centering
\begin{tabular}{|c |cccccc|}\hline
ijk$(LQ\bar{D})$   &$ 111 $ & $121$ &$ 122$ &$ 131$ 
& $123$ & $133$ 
\\ \hline 
$U(1)$ & $-11-w_1$ & 
$-6-w_1$ &  $-1-w_1$ &  $-7-w_1$ &  $-2-w_1$ 
& $-3-w_1$ \\
\hline \hline
ijk$(LQ\bar{D})$   &$ 211 $ & $221$ &$ 222$ &$ 231$ 
& $223$ & $233$ 
\\ \hline 
$U(1)$ & 
$-7-w_1$ &  $-2-w_1$ &  $3-w_1$ &  $-3-w_1$ &
$2-w_1$ & $1-w_1$ \\
\hline
\end{tabular}
\caption{{\it Integer $LL\bar{D}$ charges for Case 2: these charges remain
the same if $j$ and $k$ are interchanged.}}
\label{table:5}
\end{table}

In this model, the lepton mass matrix
takes the form

\[
V_{\ell }^{L,R}\approx \left( 
\begin{array}{ccc}
1 & \bar{\epsilon}^{2}  & 0 \\
-\bar{\epsilon}^{2}  & 1 & 0 \\ 
0 & 0 & 1
\end{array}
\right) 
\]

which is independent of the value chosen for $w_1$.
What are the predictions for the strength of the
$R$-violating couplings in this model?
As in the previous model, we have:
\bea
\lambda'_{i13} \leq 6 \cdot 10^{-4} \nonumber \\
\lambda'_{i12} \leq 6 \cdot 10^{-5} 
\eea
and so again there is no possibility to
explain the HERA events, essentially for the
same reasons as in Case 1.
From the charges of Table \ref{table:5},
we find that
$\lambda'_{211} = \lambda'_{113}$,
$\lambda'_{133} = \lambda'_{213}$ and
$\lambda'_{123} = \lambda'_{212}$.
Since  $\lambda'_{113} = \lambda'_{123}/\bar{\epsilon}
\leq 
3 \cdot 10^{-4}$, we obtain the slightly stronger bound
$\lambda'_{i13} = \lambda'_{i31} 
\leq 3 \cdot 10^{-4}$.
For the remaining couplings we have the following
bounds: \\
(i) $\lambda'_{111} = \lambda'_{112}/\bar{\epsilon} \leq 
3 \cdot 10^{-4}$, \\
(ii) $\lambda'_{122} = \lambda'_{121}/\bar{\epsilon} \leq 
3 \cdot 10^{-4}$, \\
(iii) $\lambda'_{222} = \lambda'_{221}/\bar{\epsilon} \leq 
3 \cdot 10^{-4}$, \\
(iv) $\lambda'_{223} = \lambda'_{213}/\bar{\epsilon} \leq 
      0.0013$, \\
(v) $\lambda'_{233} =  \lambda'_{223}/\bar{\epsilon} \leq 0.006$. \\
The difference from the previous solution is that
the couplings $L_3 Q_j\bar{D}_k$ are absent, and thus
constraints from them are evaded. However, the model
remains restrictive, as all the quarks of the same generation
have the same charge. Therefore, the strict bounds on products
of operators still constrain strongly individual couplings.
Nevertheless, we see from the limits above that
several possibilities exist
for $R$-violating squark decays within hadron-hadron collider detectors.

We see therefore that there are four problems
that do not allow an explanation of the HERA
events within the framework of these models.
{\bf First}, the quarks and leptons of the same
generation have the same charges, so
the $L_iL_j\bar{E}_k$ and
$L_iQ_j\bar{D}_k$ couplings are subject to the
same bounds. {\bf Secondly}, the choice of symmetric mass
matrices makes the last two equations of
(\ref{others}) difficult to satisfy, because
$Q_i$ and $\bar{D}_i$ have the same charge and
hence each factor involved has the same suppression,
so that one cannot arrange to satisfy the inequality
while keeping one coupling large. {\bf Thirdly}, the
model has large mixing in the (1,2) down-quark
sector, making the last of eqs (\ref{others})
difficult to satisfy. {\bf Finally},
the large (1,2) mixing in the charged lepton sector
may not be reconciled with bounds on
products of couplings that involve electrons
and muons.

Thus we see that the combination of the 
various $R$-violating bounds with simple 
family symmetries produces strong constraints on a 
variety of $R$-violating couplings. For the case 
of the family symmetry leading to the symmetric mass 
matrix (\ref{mm}) these constraints imply that 
$R$ violation  does not give rise to anomalous events at 
HERA at a significant rate.

\section{HERA-Friendly Textures of $R$-Violating Couplings}

In this section we explore modifications of the simple
$U(1)$ family structure, which may be able to
accommodate an $R$-violating interpretation of the apparent
excess of events at HERA.
As we have stressed, a major problem in
building a model to accommodate the HERA
events lies in the need to satisfy the bounds 
(\ref{others}) while keeping large one of the
individual couplings involved in these products.
This leads us to consider models with the (1,2)
mixing entirely in the up-quark sector and to deviate from the
symmetric mass-matrix structure.

\subsection{Asymmetric Flavour Textures}

Once one gives up on the symmetric form, the
pattern of masses is insufficient to constrain
the $U(1)$ charge structure, so there are
many new possibilities.
Here we present just one viable choice
to illustrate the options, but we
certainly do not claim any uniqueness. We start with
the charge assignment (-4,1,0) for the quark doublets
of the model discussed above, and modify the up- and
down-quark singlet charges to achieve the desired structure. In order
to reduce the arbitrariness, we
also choose to generate both the up- and
down-quark mass matrices with the same expansion 
parameter, as would be the case if the
non-renormalisable terms 
$Q_i\bar{D}_iH_2 \theta/M$
arise through heavy-quark mixing.

With this Ansatz, a suitable choice for the up-quark singlet charges is
(-5,1,0), which gives
\begin{eqnarray}
M^{up} = \left ( 
\begin{array}{ccc}
{\epsilon}^{9} & 4 {\epsilon}^{3} & {\epsilon}^4 \\ 
4 {\epsilon}^{4} & \epsilon^2 & {\epsilon} \\ 
{\epsilon}^{5} & {\epsilon} & 1
\end{array}
\right )
\end{eqnarray}
A value of $\epsilon \approx 1/20$ gives an acceptable
charm mass. Note we have been forced to
assume an enhancement factor of 4 in the (1,2) and
(2,1) elements in order to accommodate the mixing needed
to generate $V^{CKM}_{12} \approx V_{u_{12}}^L =  4 \epsilon
\approx 0.2$. 
The mass eigenvalues are $1, \epsilon^2$ and $16 \epsilon ^5$,
{} from which we see that the factor of 16 which appears from
the coefficients in the
off-diagonal entries compared to the solution
of~\cite{IR} is compensated by the additional power in the
expansion parameter.

The choice of down-quark charges is dictated by the
requirement that we keep the (1,2) mixing small.
A suitable choice for the charges of the singlet down quarks
is (7,-3,1), which gives the structure
\begin{eqnarray}
M^{down} = \left ( 
\begin{array}{ccc}
{\epsilon}^3 & {\epsilon}^{7} & {\epsilon}^3 \\ 
{\epsilon}^{8} & {\epsilon}^2 & {\epsilon}^2 \\ 
{\epsilon}^7 & {\epsilon}^3 & \epsilon
\end{array}
\right )
\end{eqnarray}
The eigenvalues scale as 
$\epsilon, \epsilon^2, \epsilon^3$, and
$V_{d_{13}}^{L} \approx 
V_{d_{31}}^{L} \approx \epsilon^2$. Moreover,
$V_{d_{23,32}}^{L} 
\approx \epsilon = 0.05$, so we do not
require the cancellations that were needed in~\cite{IR}
(remember that in this case 
$V_{d_{23,32}}^{L} \approx
\bar{\epsilon} = 0.23$).
Note that this choice has the advantage of reducing the
bottom mass through an $\epsilon$ factor, putting
us in the small-tan$\beta$ regime.
This phenomenological choice of charges does not yet ensure
anomaly cancellation, but at a later stage, when we also have a
good phenomenological choice for the lepton mass matrix, we will
discuss what flavour-independent charges have to be added
in order to cancel anomalies.
These additional charges will not modify
the hierarchy of couplings, nor the relative magnitudes 
of the $R$-violating couplings of a given type.

The key point of this model is its large $U(1)$ charge difference
between the relevant $LQ\bar{D}$ couplings:
\begin{eqnarray}
a_{L_1Q_2\bar{D}_1} - a_{L_1 Q_1 \bar{D}_2} & \rightarrow & 15 
\label{PRO1} \\
a_{L_1Q_3\bar{D}_1} - a_{L_1 Q_1 \bar{D}_3} & \rightarrow & 10 
\label{PRO2}
\end{eqnarray}

which leads to large relative suppressions of these
operators, though
mixing in the (1,2) and (1,3) down sectors will close this gap.
Consider first the operators appearing in (\ref{PRO1}). The
mixing of the left-handed down quarks is given by the form
of $V_{12,21}^L$ in the Appendix, and the second-order term
dominates with $m^d_{13} m^d_{32} / (M_3 M_2) \approx \epsilon^3 =
1.3 \cdot 10^{-4}$. The mixing of $\bar{D}_1,\bar{D}_2$ 
is given by $m_{21}^d/M_2 \approx 
m^d_{31} m^d_{23} / (M_3 M_2) \approx
\epsilon^6$.
Taking the same expansion parameter as for the
masses, consistent with (\ref{biggest}), the net suppression
is $\epsilon^3 \epsilon^6 \approx 2 \cdot 10^{-12}$. This is
more than sufficient to satisfy the
bound on $L_1Q_1\bar{D}_2$ while allowing
the $L_1 Q_2 \bar{D}_1$ to have a coefficient
large enough to give the HERA events. 
Similarly, one may check that it is possible to for the
$L_1Q_3\bar{D}_1$ operator to be relevant for HERA,
without inducing an $L_1Q_1\bar{D}_3$ coupling with an
unacceptable value.

\subsection{Lepton Flavour Violation}

We now turn to the assignment of lepton charges in such a model.
Given the bounds of (\ref{muonelectron}), if any coupling involving
an electron is large, the corresponding coupling
involving muons should be small. 
The simplest solution is to choose charge assignments
so that the $U(1)$ charge of the (1,2) entry 
of the lepton mass matrix is half-integer.
In this case, a residual $Z_2$ symmetry forbids the (1,2)
mass term. The choice of charges is restricted
by the anomaly cancellation conditions. These are:

\begin{eqnarray}
A_3 & = & ( a_{Q_1} + a_{Q_2} +a_{Q_3} ) + \frac{1}{2} (a_{\bar{U}_1} + a_{%
\bar{U}_2} + a_{\bar{U}_3}) + \frac{1}{2} (a_{\bar{D}_1} + a_{\bar{D}_2} +
a_{\bar{D}_3})  \nonumber \\
A_2 & = & \frac{3}{2} ( a_{Q_1} + a_{Q_2} +a_{Q_3}) + \frac{1}{2} (a_{L_1} +
a_{L_2} + a_{L_3} ) + \frac{1}{2} ( a_{H_1} + a_{H_2} )  \nonumber \\
A_1 & = & \frac{1}{6} ( a_{Q_1} + a_{Q_2} +a_{Q_3} ) + \frac{4}{3} (a_{\bar{U%
}_1} + a_{\bar{U}_2} + a_{\bar{U}_3}) + \frac{1}{3} (a_{\bar{D}_1} + a_{\bar{%
D}_2} + a_{\bar{D}_3})  \nonumber \\
& & + \frac{1}{2} (a_{L_1} + a_{L_2} + a_{L_3} ) + (a_{\bar{E}_1} + a_{\bar{E%
}_2} + a_{\bar{E}_3} ) + \frac{1}{2} ( a_{H_1} + a_{H_2} )  \label{eq:greens}
\end{eqnarray}

where by $a_{F_i}$ we denote the charge of particle $F$ in the $i^{th}$
generation. We see from $A_3$ and $A_2$ that, for integer quark charges, a
natural solution of the conditions has
the sums of  $a_{\bar{E}_i}$ and $a_{L_i}$ integers. Thus, we
need two of the $a_{\bar{E}_i}$  and $a_{L_i}$
to be half-integers.
We see for example, that the choice 
$a_{L_1} = 9/2, a_{L_2} = -1, a_{L_3} = -1/2,
a_{\bar{E}_1} = -1/2, a_{\bar{E}_2} = -1, 
a_{\bar{E}_3} = -1/2$
generates a viable mass hierarchy. Here we
have chosen the (3,3) charge to be the same
as that of the down-quark, in order to give
$b-\tau$ unification. Clearly this pattern of charges
is not
the only viable choice, but it indicates how things may work.
The corresponding lepton mass matrix in this solution is:

\begin{eqnarray}
M^{L} = \left ( 
\begin{array}{ccc}
{\epsilon}^4 & 0 & {\epsilon}^4\\ 
0 & {\epsilon}^2 & 0 \\ 
{\epsilon} & 0 & {\epsilon}
\end{array}
\right )
\end{eqnarray}

and there is (1,3) mixing, but no (1,2) or (2,3) mixing.
The eigenvalues of this mass matrix are
$\epsilon, \epsilon^2, \epsilon^4$, consistent with the
measured values for $\epsilon = 0.05$.

\begin{table}[h]
\centering
\begin{tabular}{|c|ccccccc|}
\hline
& $Q_i$ & $\bar{U}_i$ & $\bar{D}_i$ & $L_i$ & $\bar{E}_i$ & $H_2$ & $H_1$ \\ 
\hline
$U(1)$ & $a'_i$ & $a'_{\bar{U}_i} $ & 
$a'_{\bar{D}_i}$ & $a'_{L_i}$ & $a'_{\bar{E}_i}$ & $ 0 $ & $0 $ \\ 
\hline
$U(1)$ & $a'_i$ & $a'_{\bar{U}_i} + w_1$ & $-1+a'_{\bar{D}_i} - w_1$ & $-1-a'_{L_i}$ & $a'_{\bar{E}_i}-w_1 $ & $%
-w_1$ & $1+w_1$ \\ \hline
\end{tabular}
\caption{{\it Flavour charges in models with asymmetric mass matrices.}}
\label{table:6}
\end{table}

We denote by $a'_{Q_i} \equiv (-4,1,0)$, $a'_{\bar U_i} \equiv
(-5,1,0)$, $a'_{\bar D_i} \equiv (7,-3,1)$,
$a'_{L_i} \equiv (9/2,-1,-1/2)$ and $a'_{\bar E_i} \equiv (-1/2,-1,-1/2)$
the choices of generation-dependent charges in this model.
The top line of Table~\ref{table:6}, which includes these and the
corresponding charges for the Higgs multiplets, is not anomaly-free.
It is easy now to satisfy the anomaly-matching
conditions, allowing for additional
family-independent components of the $U(1)$ charge,
which do not affect the mass matrix structure. 
An anomaly-free solution
is obtained by adding charges as
indicated
in the second line of Table \ref{table:6}, 
where the variable $w_1$ is an integer.

\subsection{Nucleon Stability}

We now demonstrate that 
nucleon decay graphs due to
combinations of $LQ{\bar D}$ and ${\bar U}{\bar D}{\bar D}$
interactions may be eliminated in this HERA-friendly example
by imposing an anomaly-free discrete gauge symmetry.
In the specific model discussed above,
where the first- and third-generation leptons
have half-integer charges under the flavour symmetry,
we can again
combine the residual $\tilde{Z}_2$ symmetry of the $U(1)$ \cite{IR}
with a normal $Z_2^M$ matter
parity, where
$\tilde{Z}_2 \times Z_2^M$ is broken to a diagonal
$Z_2$ by a field $\Phi$ that is odd under both symmetries. Then
the couplings $\bar{U}\bar{D}\bar{D}$ and
$L_{2}Q\bar{D}$ transform as $(+,-)$ under the symmetry
and are forbidden.
Renormalisable $L_{1,3} Q \bar{D}$ couplings
which transform as $(-,-)$  are also
forbidden, but effective couplings of this type
may occur at ${\cal O}\left (\frac{\Phi}{M} \right )$
through the term $(\Phi L Q \bar{D})/M$.

Such underlying $Z_2$ symmetries can be illustrated in the context of a
GUT group, if desired.
Consider, for example, the
Pati-Salam gauge group $SU(4) \times SU(2)_L \times SU(2)_R$~\cite{PS}.
In models based on this group, the fermionic fields belong to either the
$4$ or the $\bar{4}$
representations of $SU(4)$, and no trilinear $R$-violating term is invariant
under the symmetry. However, invariants can be constructed by introducing an
adjoint field $\Sigma$~\cite{patisal}.~\footnote{A study of the 
string origin of non-renormalisable operators in this model
has been presented in~\cite{string}.} If $\Phi$ has a half-integer
charge,
the fact that it is in the adjoint of $SU(4)$ means that all
baryon-number-violating operators are forbidden in any order, whilst the
terms $L_{1,3}Q\bar{D}\Phi$ have integer charge and are therefore allowed.
Moreover, no effective terms $L_{1,3}\bar{E}H_2\Phi$, which could cause
problems with the lepton mass hierarchies, are allowed, as they are not
invariant under the extended gauge group.

\subsection{Hierarchy of $R$-Violating Interactions in 
a HERA-Friendly Model}

We now consider the effect of the $U(1)$ symmetry on the pattern 
of allowed $R$-violating interactions in the models that were
motivated by the HERA events. The charges of the operators depend on the
half-integer charge of the field $\Phi$ under the flavour symmetry. This
does not affect the relative magnitudes of the $R$-violating couplings,
since $\Phi$ appears in all terms. However, this charge and the vacuum
expectation value of $\Phi$ do provide a possible source of suppression for
the $R$-violating couplings. 
The corresponding subclasses of $LL\bar{E}\Phi$ and $LQ\bar{D}\Phi$
operators with integer flavour charge, before introducing mixing effects,
appear in Tables~\ref{table:7} and \ref{table:8}. We 
have used the charges of Table~\ref{table:6},
imposing anomaly cancellation. Here we
have taken $\Phi$ to have $U(1)$ charge $1 / 2$, but its actual value
can be re-absorbed in the definition of $w_1$.

Let us first consider the possibility that the 
possible excess HERA events are due to the
$L_1Q_3\bar{D_2}$
operator with $\lambda^{\prime}> 0.3 / \sqrt{B}$, 
which will be the dominant
operator if $w_1 = 0$. The relative suppression of the $L_1 Q_3 \bar{D}_3$ 
operator is $\epsilon^4$.
Of course, mixing effects in the
(2,3) sector of the down-quark mass matrix 
re-introduce an $L_1Q_3\bar{D_3}$ operator
at order $\epsilon^2$, via the mixing of right-handed down quarks.
In the type of models with a small $V^R_{d_{2,3}}$ mixing 
therefore, this solution may in principle be accommodated.
However, for our specific choice of charges, we see that the
operator $L_1 Q_3 \bar{D}_2$ has the same charge as
$L_1 Q_1 \bar{D}_3$, which is bound 
by charged-current universality~\cite{noi} to be $\leq 0.04$
for a squark mass of 200 GeV.
Hence the $L_1Q_3\bar{D}_2$ interpretation of
the HERA data is not realisable in this
specific model. However, this is rather 
accidental for this particular example, and need
not be the case in general.

\begin{table}[h]
\centering
\begin{tabular}{|c|cccc|}
\hline
ijk$(LL\bar{E}\Phi)$ & $122 $ & $131$ & $133$ & $231$ \\ 
\hline
$U(1)$ & $1-w_1$ & $2-w_1$ & $2-w_1$ & $-4-w_1$ \\ 
\hline
\end{tabular}
\caption[]{\it Integer $LL\bar{E}\Phi$ charges, ignoring
mass mixing.}
\label{table:7}
\end{table}

\par
\vspace*{0.8 cm}
\par
\begin{table}[h]
\centering
\begin{tabular}{|c|cccccc|}
\hline
ijk$(LQ\bar{D}\Phi)$ & $111 $ & $112$ & $113$ & $121$ & $122$ & $123$ \\ 
\hline
$U(1)$ & $6-w_1$ & $-4-w_1$ & $-w_1$ & $11-w_1$ & $1-w_1$ & $5-w_1$ \\ 
\hline\hline
ijk$(LQ\bar{D}\Phi)$ & $131 $ & $132$ & $133$ & $311$ & $312$ & $313$ \\ 
\hline
$U(1)$ & $10-w_1$ & $-w_1$ & $4-w_1$ & $1-w_1$ & $-9-w_1$ & $-5-w_1$ \\ 
\hline\hline
ijk$(LQ\bar{D}\Phi)$ & $321 $ & $322$ & $323$ & $331$ & $332$ & $333$ \\ 
\hline
$U(1)$ & $6-w_1$ & $-4-w_1$ & $-w_1$ & $5-w_1$ & $-5-w_1$ & $-1-w_1$ \\ 
\hline
\end{tabular}
\caption{{\it Integer \protect$LQ\bar{D}\Phi\protect$ charges, ignoring
mass
mixing.}}
\label{table:8}
\end{table}

We now look at the
possibilities that the HERA events arise from the $L_1Q_{2,3}\bar{D}_1$
operators. We see from Table~\ref{table:6} that, 
in the absence of mixing, the relative
suppressions of the $L_1Q_1\bar{D}_1$ and $L_1Q_3\bar{D}_3$ operators would
have been enough to make these cases viable. When mixing effects are
included, the possible effects of unknown phases should be taken into
account when comparing with bounds.
In the specific case that the HERA events are due to an $L_1Q_2\bar{D}_1$
coupling, we have no problem with the $L_1Q_3\bar{D}_3$ operator, but there
is a potential difficulty with
$\beta \beta$ decay, due to mixing with the
$L_1Q_1\bar{D}_1$ operator:

\begin{equation}
\left| \lambda_{112} V^L_{u_{12}}  \left(\frac{200 \; GeV}{m_{\tilde{u}_L}}%
\right)^{\!\!2} \right|<4\cdot 10^{-3} \left(\frac{m_{\tilde{g}}}{1 \; TeV}%
\right)^{\!1/2}
\end{equation}

where $m_{\tilde{g}}$ is the gluino mass. Given (a) that the $V^{CKM}$
mixing arises from the up sector in our framework~\footnote{
Even in the case that the $V^{CKM}_{12,21}$ mixing
arises from the down-quark sector,
squark mass universality violation is required in
order to evade bounds from
$K \rightarrow \pi \nu \bar{\nu}$.
}, and (b) that the bounds
from the Tevatron indicate that the branching ratio of $\tilde{c}_L$ to
fermions can not be close to unity in the context of this interpretation,
implying that $\lambda^{\prime}_{121}$ has to be larger
than 0.04, we see that this solution is not naturally accommodated.
It might be possible if $m_{\tilde{u}_L}$ is significantly larger than
$m_{\tilde{c}_L}$, but this
requires a violation of squark-mass universality that is
potentially dangerous for flavour-changing neutral interactions.

On the other hand, if
the HERA events are due to a $L_1Q_3\bar{D}_1$ coupling, there is no
problem with the $L_1Q_1\bar{D}_1$ operator, but a problem could
in principle appear with the $L_1Q_3\bar{D}_3$ coupling
that is bounded from limits on neutrino Majorana masses~\cite{numass}.
However, the
relevant (3,1)
mixing term is small, indicating that in
this case the unknown coefficients may be such that the bounds are easily
accommodated. What about the other couplings? 
For $L_1Q_3\bar{D}_1 \approx 0.04$, the model predicts that
$L_1Q_2\bar{D}_1 \approx 
L_3Q_1\bar{D}_2 \approx 
0.002$, while all other couplings are very
suppressed. Indeed, looking at the charges, we see
that the next larger couplings are
suppressed by $\epsilon^4$ as compared to
$L_1Q_3\bar{D}_1$. Mixing effects 
are also suppressed, except for the
operators 
$L_1Q_1\bar{D}_1 \approx  L_1Q_2\bar{D}_1 (4\epsilon) 
\approx 0.0004$
and
$L_3Q_2\bar{D}_2 \approx  L_3Q_1\bar{D}_2 (4\epsilon) 
\approx 0.0004$,
 which are within the allowed range.

Finally, note that we do not have any
mixing between $L_1Q_j\bar{D}_k$ and $L_2Q_j\bar{D}_k$ 
couplings (the later are forbidden by the symmetry),
so the dangerous product combinations
that violate lepton flavour
are also absent. 

In the light of the above discussion, we conclude that,
of the valence quark production mechanisms via $L_1Q_2\bar{D}_1$ and $%
L_1Q_3\bar{D}_1$ couplings, the second possibility seems to be
favoured. It should be possible to make
a model with a coupling $L_1Q_3\bar{D}_2$ sufficiently large
to explain the HERA data, although we have not displayed one here.

\section{Baryon Decay via Dimension-Five Operators}

We saw earlier on that the experimental absence of baryon decay imposed
important constraints on possible models, which are most easily evaded by
imposing a baryon parity symmetry that forbids the dangerous ${\bar U} {\bar
D} {\bar D}$ couplings. However, this is not the end of the story, since
models may also
contain dimension-five operators that would generate proton decay
at an unacceptable level. The most dangerous among these operators are the
operators $[QQQL]_F$ and $[QQQH_1]_F$, the latter in the presence of
$LQ\bar{D}$ couplings.~\footnote{The lepton-number-violating 
operators $[Q\bar{U}\bar{E}H_1]_F$, $[Q\bar{U}
L^*]_D$ and $[Q\bar{U}L^*]_D$ are dangerous in the presence of $\bar{U}\bar{D%
}\bar{D}$ ones.} These operators can lead to fast proton decay via loop
diagrams. In the case of $[QQQL]_F$ operators that involve the two lightest
generations, the constraint on the coupling $\eta$ of any such operator is 
$\eta \le 10^{-7}$. This bound has some flexibility, since the magnitude of
the loop diagrams depends on details of the sparticle spectrum, but this
possibility is not crucial for the subsequent discussion of models. In the
case of $[QQQH_1]_F$ operators with couplings $\eta^{\prime}$, fast proton
decay may occur if they are present simultaneously with $LQ\bar{D}$
operators with generic coefficients $\lambda^{\prime}$. The product of the
corresponding couplings is constrained: $\eta^{\prime}\lambda^{\prime}\le
10^{-10}$. Since an $R$-violating interpretation of the HERA events requires
either $L_1Q_2\bar{D}_1$ or $L_1Q_3\bar{D}_1$ $\approx 0.04$, it is clear
that we have to worry about the $[QQQH_1]_F$ operator as well.

We now analyse the dimension-five operator charges in the different cases
discussed in previous sections, to see whether they are large enough for the
suppression by powers of small quantities to be sufficient. How small the
terms actually are depends on the expansion parameter, as we have already
discussed in a previous section.

The $QQQH_1$ operators are easily dealt with, even
though the baryon stability
requirements seem to be more severe for them. The reason is that these
operators
transform as $(+,-)$ under the $\tilde{Z}_2 \times Z_2^M$, and are thus
forbidden.
What about the $QQQL$ operators? The $QQQL_{1,3}$ operators are not present,
because they transform as $(-,+)$ under $\tilde{Z}_2 \times Z_2^M$. However,
the operators $QQQL_2$ are allowed. These are dangerous, because proton
decay may occur via the modes $p \rightarrow \bar{\nu}_{1,2,3}
\pi^+$ and $p \rightarrow \bar{\nu}_{1,2,3}K^+$.
Let us look at the flavour charges of these operators. We recall that colour
antisymmetrisation implies that all the quark flavour indices cannot be
identical. The operators that are not suppressed enough by quark mixing
parameters have the following charges in the model that could
explain the HERA events:
\begin{eqnarray}
a_{Q_1Q_1Q_2L_2} & = & -9  \nonumber \\
a_{Q_1Q_2Q_2L_2} & = & -4  \nonumber \\
a_{Q_1Q_1Q_3L_2} & = & -10  \nonumber \\
a_{Q_1Q_2Q_3L_2} & = & -5
\end{eqnarray}

where for the lepton charge we used the anomaly-free choice of
Table~\ref{table:6}.

We infer that we do not
need any further underlying symmetry in order to suppress these
couplings adequately. However, even in models where this 
suppression does not occur,
there could be some GUT symmetry that
forbids the offending $QQQL$ operators~\footnote{
Moreover, in string-derived GUT models, string selection
rules may lead to the vanishing of 
operators that are invariant under
field-theory symmetries. In such models, it is possible to
construct realistic fermion mass matrices while having
maximal proton stability \cite{ELLN}.}.
This 
would be an interesting constraint on
GUT model-building, but should not be taken as a serious obstacle to
constructing HERA-friendly models.

\section{Concluding Comments}

We have discussed the implications of a single $U(1)$ abelian flavour
symmetry for the possible hierarchies of $R$-violating couplings. The
relations between the Standard-Model Yukawa couplings and $R$-violating
couplings depend on the choice of model charges, so the observed hierarchies
of quark and lepton masses do not lead to a unique specification of the
dominant $R$-violating couplings. However, we have identified
certain general features of such a framework, highlighting the
importance of mass mixing between current eigenstates.
We have identified various interesting possibilities for
hadron-hadron collider phenomenology that are consistent
with this mixing and the available experimental constraints.
Within this general approach, we have searched 
specifically for simple consistent models
that lead to the favoured $R$-violating scenarios for the 
explaining the possible excess in the HERA data.

Our results may be summarised as follows:

$\bullet$ Flavour symmetries lead us to expect a hierarchy in the 
$R$-violating couplings, analogous to that observed for the known
fermion masses. These hierarchies can be consistent with 
a squark-production interpretation of the HERA data (if required), as
well as with the various other experimental constraints on the couplings.

$\bullet$ The simplest charge assignments lead to unified,
and
thus more predictive, forms for the 
mass matrices. For the case of equal charges 
for up and down quarks and leptons of a given 
generation, the symmetry together with bounds from products of
$R$-violating couplings implies that there 
should be no significant anomalous  events 
at HERA coming from such couplings. If we wish to accommodate such anomalous events,
 we are forced to depart from this picture. 
Schemes with asymmetric charges and different assignments for up
quarks, down quarks and leptons give rise to larger splittings
between different operators.

$\bullet$ Some of the charge assignments considered forbid
large coefficients
of dimension-five operators that are potentially dangerous for baryon
stability.  In schemes where this is not true,
such terms would need to be forbidden by further GUT symmetries.

One can consider relaxing various of our conditions, for example 
by introducing
a higher level of asymmetry in the mass matrices, invoking multiple $U(1)$
flavour symmetries, etc., and in such models the predictions can be further
altered. Moreover, additional zero couplings may be expected when one goes
to a specific GUT/string construction. However, it is interesting that it is possible to construct phenomenological models with a single $U(1)$ flavour symmetry that are compatible with attempts to
explain the reported excess of HERA data by $R$-violating squark
production, albeit at a price. In order to constrain the possible schemes, and perhaps rule some out, more experimental data are required. 

\vspace*{0.2 cm}

\begin{center}

{\bf Appendix}

\end{center}
Using second-order perturbation theory, it is easy to derive the mixing
elements for a generic mass matrix~\cite{HR,BJ},
where $m$ stands for the off-diagonal
contributions and $M$ for the diagonal part.
The left-handed mixing is given by \cite{BJ}
\bea
V_{ij}^L = - \frac{
(m_{ij}M_j+m^*_{ji}M_i)}{M_i^2-M_j^2} +
\frac{ (m_{ik}M_k+ m_{ki}^*M_i)
       (m_{kj}M_j +m^*_{jk} M_k ) }
     {(M_i^2-M_j^2)(M_k^2-M_j^2)} -
\frac{m_{ik} m^*_{jk}}{(M_i^2-M_j^2)} \nonumber
\eea
and $V_{ij}^R$ is given by a
corresponding expression, substituting the
mass matrix by its hermitian conjugate.
In the case that the ratio of
$m_{ij}$ to $m_{ji}^*$ is
considerably larger than the ratio $M_i/M_j$,
the mixing elements are given by:
\begin{eqnarray}
V^L_{12} & = & +\frac{m_{12}}{M_2} \; - \left [ \frac{m_{13}m_{32}}{M_3 M_2}
\right ] \; + \; ...  \nonumber \\
V^L_{21} & = & -\frac{m^*_{12}}{M_2} \; + \left [ \frac{m^*_{32}m^*_{13}}{M_2
M_3} \right ] \; + \; ...  \nonumber \\
V^L_{13} & = & +\frac{m_{13}}{M_3} \; + \left ( \frac{m_{12}m_{23}M_2}{M_3^3}
\right) \; + \left [ \frac{m_{12}m^*_{32}}{M_3^2} \right ] \; + \; ... 
\nonumber \\
V^L_{31} & = & -\frac{m_{13}^*}{M_3} \; + \frac{m^*_{12}m^*_{23}}{M_2 M_3} \;
+ \left[ - \frac{m_{32}m^*_{12}}{M_3^2} \right ] \; + \; ... \\
V^L_{23} & = & +\frac{m_{23}}{M_3} \; + \left ( \frac{m^*_{12}m_{13}M_2}{M_3^3}
\right) \; + \left [ \frac{m_{21}m^{*}_{31}}{M_3^2} \right ] \; + \; ... 
\nonumber \\
V^L_{32} & = & -\frac{m^*_{23}}{M_3} \; - \frac{m_{12}m^*_{13}}{M_2 M_3} \; + 
\left [ - \frac{m_{31}m^*_{21}}{M_3^2} \right ] \; + \; ...  \nonumber
\label{mix}
\end{eqnarray}

In the above expressions, 
terms in brackets mark contributions which
involve mass entries below the diagonal. These terms,
as well as the ones in parentheses, can in most
cases be neglected. However, they
may become relevant if the mass matrices have texture
zeroes.

In the case of the full $V_{CKM}$ matrix, one has

\begin{eqnarray}
V^{CKM}_{12} & = & + \frac{m^d_{12}}{M_2^d} - \frac{m^u_{12}}{M_2^u} + ... 
\nonumber \\
V^{CKM}_{21} & = & - \frac{{m^d_{12}}^*}{M_2^d} + \frac{{m^u_{12}}^*}{M_2^u}
+ ...  \nonumber \\
V^{CKM}_{23} & = & + \frac{m^d_{23}}{M_3^d} - \frac{m^u_{23}}{M_3^u} + \frac{%
m^d_{13} {m_{12}^u}^*}{M_3^d M_2^u} - \frac{m^u_{13} {m_{12}^u }^*}{M_3^u
M_2^u} + ...  \nonumber \\
V^{CKM}_{32} & = & -\frac{{m^d_{23}}^*}{M_3^d} + \frac{{m^u_{23}}^*}{M_3^u}
- \frac{m^d_{12} {m_{13}^d}^*}{M_2^d M_3^d} + \frac{m^d_{12} {m_{13}^u}^*}{%
M_3^u M_2^d} + ...  \nonumber \\
V^{CKM}_{13} & = & + \frac{m^d_{13}}{M_3^d} - \frac{m^u_{13}}{M_3^u} - \frac{%
m^d_{23} {m_{12}^u}}{M_3^d M_2^u} + \frac{m^u_{12} {m_{23}^u}}{M_3^u M_2^u}
+ ...  \nonumber \\
V^{CKM}_{31} & = & -\frac{{m^d_{13}}^*}{M_3^d} + \frac{{m^u_{13}}^*}{M_3^u}
+ \frac{{m^d_{12}}^*{m_{23}^d}^*}{M_2^d M_3^d} - \frac{{m^d_{12}}^*{m_{23}^u}%
^*}{M_3^u M_2^d} + ...  \label{secmix}
\end{eqnarray}

These formulae are used in the text in conjunction with specific parametric
forms for the off-diagonal terms in the up- and down-quark mass matrices.

\end{document}